\documentclass[11pt]{article}

\usepackage[utf8]{inputenc}
\usepackage[T1]{fontenc}
\usepackage{lmodern}
\usepackage[margin=1in]{geometry}
\usepackage{setspace}
\usepackage{booktabs}
\usepackage{longtable}
\usepackage{array}
\usepackage{tabularx}
\usepackage{placeins}
\usepackage{hyperref}
\usepackage{csquotes}
\usepackage{enumitem}
\usepackage[round]{natbib}

\newcommand{\pdfauthormeta}{Ishrat Jahan Easha; Nabil Mosharraf Hossain; Araf Mohammad Mahbub; Fairoze Bint Abu Hassan; Zunaid Aslam; Yemin Sajid; Riasat Islam}

\newcolumntype{Y}{>{\raggedright\arraybackslash}X}

\newcommand{\manuscripttitle}{Designing for Interconnected Islamic Learning: A Qualitative Study of Muslim Women's Experiences with Qur'an, Hadith, and Seerah Apps}
\newcommand{\manuscriptkeywords}{digital religion; Islamic apps; human--computer interaction; interconnected learning; epistemic trust}

\hypersetup{
  colorlinks=true,
  linkcolor=black,
  citecolor=black,
  urlcolor=blue,
  pdftitle={\manuscripttitle},
  pdfsubject={Submission manuscript for the International Journal of Human-Computer Interaction},
  pdfkeywords={\manuscriptkeywords},
  pdfauthor={\pdfauthormeta}
}

\title{Designing for Interconnected Islamic Learning:\\
A Qualitative Study of Muslim Women's Experiences with Qur'an, Hadith, and Seerah Apps}

\author{Ishrat Jahan Easha$^{1,2}$, Nabil Mosharraf Hossain$^{3}$, Araf Mohammad Mahbub$^{3}$,\\
Fairoze Bint Abu Hassan$^{3}$, Zunaid Aslam$^{3}$, Yemin Sajid$^{3}$, Riasat Islam$^{3,4}$\thanks{Corresponding author: riasat.islam@qmul.ac.uk}\\[0.5em]
\small $^1$University of Technology Sydney\\
\small $^2$ZNRF University of Management Sciences\\
\small $^3$Greentech Apps Foundation\\
\small $^4$Queen Mary University of London}
\date{}

\begin{document}
\maketitle

\begin{abstract}
Islamic learning often depends on reading the Qur'an, Hadith, and Seerah together, yet digital tools typically separate these sources across apps, screens, and search pathways. We examine this as a human--computer interaction problem through five semi-structured interviews with Muslim women recruited from an online Islamic learning community. Participants described a recurring tension: they wanted Qur'an--Hadith--Seerah context at the point of reading, but only when contextual expansion remained trustworthy, optional, and did not interrupt reading. Interpreting the interviews through gendered digital religion, epistemic trust, and seamless learning, we identify five themes concerning contextual understanding, authenticity, interface clutter, study modes, and guidance features. We introduce layered contextuality as an HCI account of this domain: contextual expansion must be balanced with interpretive accountability, devotional flow, and continuity across devices and study intensities.
\end{abstract}

\textbf{Keywords:} \manuscriptkeywords

\section{Introduction}

Digital religious practice has become a major site of contemporary interaction design. Research on digital religion has shown that digital media do not simply reproduce offline religious life; they reshape authority, practice, access, and community in consequential ways \citep{campbell2013digital,bunt2009imuslims,solahudin2019internet}. Within Islamic digital ecosystems, users routinely encounter separate applications or websites for Qur'an recitation and English renderings, hadith collections, supplications, Seerah, tafsir, prayer reminders, and memorisation support. This abundance creates access, but it also creates fragmentation. A learner may read a verse in one app, search for explanatory hadith elsewhere, and then move to a different resource to understand the historical event or Prophetic biography related to that teaching. Prior work on Islamic apps already suggests that users value spiritually meaningful and trustworthy support, but it also indicates that many systems remain organised around discrete functions rather than integrated learning journeys \citep{kabir2025islamiclifestyle,mubin2020reviewing}.

This fragmentation matters because classical Islamic learning is rarely structured as isolated text consumption. The Qur'an, Hadith, and Seerah are commonly read in relation to each other to establish context, interpretive boundaries, and practical application. Islamic scholarship on contextual interpretation similarly emphasises that meaning emerges through relations among text, historical occasion, Prophetic example, and commentary rather than through decontextualised extraction alone \citep{saeed2021contextualist,calis2022contextual,ramle2022between}. When digital tools separate these sources, users may gain convenience while losing coherence. From an HCI perspective, this creates a design problem that is only partly about content management. It is also a problem of navigation, cognitive load, trust, progressive disclosure, and support for different learning modes.

Here we examine that problem through a qualitative interview study of Muslim women's experiences with Islamic apps. We centre Muslim women because prior work on Islamic digital platforms has tended to aggregate Muslim users more broadly, leaving women's situated learning practices comparatively underexamined as a primary analytic focus \citep{kabir2025islamiclifestyle,midden2013digitalfaiths,nisa2022muslims}. Recent HCI work on Islamic lifestyle applications has already shown that Muslim users want trust, understanding, and spiritually meaningful digital support, but it has not isolated women's perspectives as a primary analytic lens \citep{kabir2025islamiclifestyle}. Drawing on work in gendered digital religion, we treat these accounts as situated knowledge about how religious authority, credibility, access, and everyday practice are negotiated through media \citep{lovheim2013media,nas2022women}. We ask three research questions:

\begin{enumerate}[leftmargin=1.5em]
  \item How do Muslim women currently use Qur'an, Hadith, and Seerah apps in everyday religious learning?
  \item How do they conceptualise \enquote{interconnected learning} across these sources?
  \item What interaction and feature patterns do they see as helpful, disruptive, or missing?
\end{enumerate}

Using semi-structured interviews and thematic analysis, we identify five themes that support one main HCI argument. Participants did not describe interconnected learning as merely linking sources across texts. They asked for context anchored to the immediate reading moment, deeper material available on demand, provenance visible enough to support trust, and continuity between devotional mobile use and more intensive study. We refer to this pattern as layered contextuality and present it as an HCI contribution centred on Muslim women to the design of Islamic learning technologies. Gendered digital religion, epistemic trust, and seamless learning guide our interpretation because they captured three recurrent dimensions of the dataset: situated routines, interpretive accountability, and movement across devices and study intensities \citep{lovheim2013media,nas2022women,fogg1999credibility,fogg2003prominence,corritore2005mechanics,kearney2020seamless,brudy2019crossdevice}.

\section{Related Work}

\subsection{Digital religion and Islamic digital practice}

Digital religion scholarship has shown that digital media reshape how religious authority, practice, and belonging are negotiated \citep{campbell2013digital,bunt2009imuslims}. In Islamic contexts, internet and social media environments have already transformed learning practices, authority formation, and access to religious knowledge \citep{solahudin2019internet}. These shifts are not neutral: they alter how users find sources, who is trusted, what counts as legitimate interpretation, and how individuals move between personal devotion and public knowledge.

At the same time, digital religious tools do not necessarily produce integrated learning experiences. Research on mobile religious practice has shown that apps can support devotional repetition, remembrance, and access, while still reflecting the market, technical, or interaction constraints of contemporary platforms \citep{sijapati2019sufi}. In Islamic app ecosystems, this often means that a user encounters many narrow tools rather than one system that supports contextualised learning across genres and sources. Mubin et al.'s review of Qur'an learning apps highlights how existing tools often emphasise discrete learning functions rather than integrated interpretive support \citep{mubin2020reviewing}.

\subsection{Centring Muslim women's voices in digital religion research}

Work on Muslim women and digital media has shown that online spaces are important sites of agency, identity work, and religious expression, while also remaining shaped by unequal gendered power relations \citep{midden2013digitalfaiths,warren2018placingfaith,nisa2022muslims}. L\"ovheim argues that gender has too often remained an \enquote{add-on} in media and religion research rather than a core analytic lens \citep{lovheim2013media}. This matters here because accounts centred on Muslim women can reveal design needs that generalised user studies flatten or miss. Recent HCI work on Islamic lifestyle applications similarly shows that Muslim users want guidance, trust, flexibility, and spiritually meaningful forms of digital support, but it does not isolate women's perspectives as a primary analytic frame \citep{kabir2025islamiclifestyle}. Nas's study of the \enquote{Women in Mosques} campaign adds a complementary point: digital media can make gendered religious exclusions visible by circulating women's testimonies \citep{nas2022women}. Although our study examines learning apps rather than mosque activism, the connection is analytically useful because in both cases mediated accounts reveal how access to religious participation is structured through design, visibility, and authority.

This gap has also been named directly within HCI. Mustafa et al. argue that Muslim women's needs and agendas remain underrepresented in HCI discourse and design, particularly when dominant design lenses sideline Islamic values and everyday faith practices \citep{mustafa2020islamichci}. Ibtasam et al. show that women's technological inclusion in Islamic culture is mediated by family relations and gendered norms, shaping access to phones, mobile shops, and digital participation \citep{ibtasam2019family}. Conversely, Rifat et al.'s study of Islamic religious communities reports that restricted access to women in madrasahs and mosques produced a participant group heavily shaped by male participation, underscoring how easily Muslim women's perspectives can be missed in HCI studies of Muslim communities \citep{rifat2020religion}.

\subsection{HCI perspectives on values, credibility, and culturally aware design}

We situate this study within HCI conversations about values, credibility, and culturally specific design. Feminist HCI argues that design should engage pluralism, participation, advocacy, and embodiment rather than assume a universal user \citep{bardzell2010feminist,bardzell2011feminist}. Reflective HCI similarly asks designers to expose the assumptions built into systems and to treat the making of meaning as a legitimate interactional concern \citep{sengers2006reflective}. In parallel, computer credibility research has shown that trust depends not only on information quality, but on how systems signal reliability, prominence, and interpretability \citep{fogg1999credibility,fogg2003prominence,corritore2005mechanics}. Recent HCI work has extended these concerns into explicitly cultural and religious domains: Mustafa et al. argue that design with Muslim populations must engage religious practices on their own terms \citep{mustafa2020islamichci}; work on cultural sensitivity highlights how technologies participate in negotiating visibility and norms \citep{hakkila2020cultural,alshehri2022culturally}; and research on religious and spiritual artefacts shows that sacred practices are legitimate HCI design domains \citep{markum2023religious}. We build on this literature by showing that Muslim women's accounts reshape the design problem of interconnected religious learning into a question of contextual interpretation, credibility, and continuity across contexts.

\subsection{A scholarly perspective on interconnected Islamic learning}

Interconnected learning is not simply a product idea. It also reflects a longstanding scholarly concern with how Islamic texts are interpreted in relation to report, context, and Prophetic example. Saeed and Akbar show that contextualist Qur'anic interpretation foregrounds historical setting, audience, and ethical purpose \citep{saeed2021contextualist}. Related work on Qur'anic contextualisation and on \enquote{asb\={a}b al-wur\={u}d} in hadith studies similarly emphasises that isolated textual fragments are often insufficient for interpretation without knowledge of circumstance, narrative setting, and scholarly explanation \citep{calis2022contextual,ramle2022between}. Across these debates, context is not an optional supplement but a condition of understanding.

For HCI, the key point is not to adjudicate theological disagreement. The relevant point is that participants' desire to move between verse, hadith, commentary, and Prophetic biography aligns with an established scholarly concern for situated understanding. Framing the problem this way helps us position interconnected learning as more than feature integration. It is a design response to a scholarly and pedagogical logic in which meaning often emerges through relationships among texts, interpretation, and context. This perspective also complements L\"ovheim's call to analyse how media, religion, and gender interact in the making of religious meaning and authority rather than treating gender as a peripheral variable \citep{lovheim2013media}.

\subsection{Interconnected learning as a seamless learning problem}

The notion of interconnected Islamic learning is closely related to work on seamless learning: learning that crosses contexts, devices, moments, and knowledge representations rather than remaining trapped in a single formal environment \citep{kearney2020seamless}. In our study, participants moved between devotional reading, search, reflection, memorisation, and research writing. Their accounts suggest that \enquote{interconnection} is not just about linking content objects. It is about helping learners maintain continuity across tasks, contexts, and interpretive layers. This interpretation also aligns with HCI work on cross-device interaction, which treats continuity across multiple surfaces as a distinct design challenge rather than an implementation detail \citep{brudy2019crossdevice}.

This makes interface quality especially important. Mobile learning research has shown that users are sensitive to clutter, usability, and friction in app environments \citep{okuboyejo2021concerns}. Work on self-regulated learning with digital tools similarly suggests that prompts, planning aids, and feedback are valuable only when they support rather than interrupt meaningful engagement \citep{wong2019selfregulated,kori2014reflection}. In religious learning, that sensitivity is heightened because many tasks are voluntary, reflective, or spiritually motivated. Minor disruptions may therefore carry disproportionate cost.

\subsection{Islamic education, access, and user diversity}

Educational technology research in Islamic contexts has drawn attention to uneven access, differing user expectations, and the importance of matching tools to actual study practices \citep{hardaker2017perceptions}. Existing HCI work on Islamic lifestyle apps similarly shows that Muslim users value guidance, flexibility, and spiritual relevance, but it also points to substantial diversity in how people want such systems to fit everyday life \citep{kabir2025islamiclifestyle}. We extend that concern by showing that even within a relatively committed user group, there are sharply different modes of use. Some participants described mobile apps as portable devotional companions; others treated them as secondary tools compared with working on the web or using Arabic research workflows. That diversity matters for HCI because it suggests that a single design logic is unlikely to serve all religious learners equally well.

\section{Method}

\subsection{Research design and interview protocol}

We used a qualitative design based on semi-structured interviews. Our aim was to understand how Muslim women currently use Islamic apps, how they conceptualise interconnected learning across Qur'an, Hadith, and Seerah, and what kinds of interface features they see as supportive or disruptive in that process. Semi-structured interviewing was appropriate because we sought situated accounts of everyday religious learning, practices of switching between apps, trust judgments, and design expectations rather than standardised measures \citep{kallio2016semistructured}.

We organised the interview protocol around three areas: participants' current use of Islamic apps and digital study routines; their understanding of contextual or interconnected learning across Islamic sources; and their responses to sensitising design probes grounded in literature about how contextual links, inspectable explanation, and learning support might be represented in an interface, including embedded links, expandable commentary or provenance cards, planners, quizzes, and reflection prompts. This structure allowed us to move from existing practice to prospective design, which was important for an HCI study concerned with both lived experience and implications for interaction design. It also allowed participants to define for themselves what counted as meaningful, trustworthy, or disruptive support in a religious learning interface.

These probes were not presented as preferred solutions. They were derived from four concerns already established in adjacent work and from feature patterns visible in Islamic app ecologies. First, Islamic learning scholarship and app research both point to the importance of linking Qur'an, hadith, Seerah, and commentary for contextual understanding rather than treating them as isolated resources \citep{mubin2020reviewing,kabir2025islamiclifestyle,saeed2021contextualist,calis2022contextual,ramle2022between}. Second, HCI and digital religion research suggest that credibility depends on inspectable provenance, explanatory cues, and the ability to judge why a connection can be trusted \citep{fogg1999credibility,fogg2003prominence,corritore2005mechanics,solahudin2019internet}. Third, seamless learning and cross-device HCI motivate attention to continuity across devices, contexts, and study intensities \citep{kearney2020seamless,brudy2019crossdevice}. Fourth, educational research on self-regulated and reflective learning supports the use of optional prompts, planning aids, and feedback that guide engagement without overwhelming it \citep{okuboyejo2021concerns,wong2019selfregulated,kori2014reflection}. We therefore used these prompts as sensitising probes to elicit participants' judgments about what kinds of contextual support would feel useful, trustworthy, or disruptive in a religious learning interface.

\subsection{Participants and recruitment}

We recruited participants from a Muslim women's WhatsApp group associated with an online course in foundational Islamic studies. Eligible participants were Muslim women aged 18 or above, including both teachers and students at different levels of study. Recruitment was voluntary, no incentives were offered, and participation took place through online interviews. Our final sample comprised five participants. We approached sample adequacy through information power rather than numerical saturation: given the specificity of the participant group, the focused research questions, and the analysis informed by theory, five interviews provided a corpus rich in information for this focused qualitative HCI study \citep{malterud2016information,braun2021saturation,crabtree2025human}.

The sample was socially and educationally specific. P1 was based in Bangladesh, while P2--P5 were based in the United Kingdom. Participants were in their 30s and 40s, all had postgraduate education in their own fields, and all were regular users of smartphones and Islamic digital resources. Their Islamic study backgrounds varied from learners connected to the course to users focused on teaching and more advanced users focused on study who described using apps alongside lectures, websites, or more formal study materials. All participants were Sunni Muslims. In aggregate, one participant was a housewife, one was a full-time student, and three were in full-time employment. For the purposes of this study, denominational affiliation and employment status were background descriptors rather than focal analytic variables, and we therefore report them only in aggregate.

To protect confidentiality, we report participants as P1--P5. Table~\ref{tab:participants} summarises descriptive participant background together with concise study and app use descriptors. We keep these descriptors close to reported practice so that readers can distinguish factual background from later interpretation.

\begin{table}[ht]
\centering
\caption{Participant background and descriptive study/use profiles}
\label{tab:participants}
\small
\begin{tabularx}{\textwidth}{p{0.9cm}p{2.0cm}p{2.7cm}Y}
\toprule
ID & Location & Reported study and app use pattern & Salient interview details \\
\midrule
P1 & Bangladesh & Frequent app use, including use related to teaching & Emphasised authenticity, integrated access, bookmarks, planners, and memorisation support. \\
P2 & UK & Primarily mobile devotional and learning use & Framed interconnected learning as holistic understanding; wanted a seamless master app, goals, and quizzes. \\
P3 & UK & Mixed devotional and study focused use & Emphasised trust, contextual understanding, optional links, planners, and everyday itineraries. \\
P4 & UK & Selective app use focused on prayer and core functions & Saw value in interconnected content but warned against clutter, feature overload, and disruptive ads. \\
P5 & UK & Intensive study across app and web resources & Valued English glosses, bookmarks, reflection, search, commentary, copying and pasting, and website/laptop workflows. \\
\bottomrule
\end{tabularx}
\end{table}

\subsection{Data collection}

We collected data through semi-structured online interviews lasting approximately 30--45 minutes. In this study, a female interviewer conducted the sessions as a culturally sensitive methodological choice that supported participant comfort, direct access, and fuller engagement with a cohort that HCI work has often underrepresented or reached unevenly \citep{mustafa2020islamichci,rifat2020religion}. The interviews invited participants to describe which Islamic apps they used, how often and for what purposes they used them, how they moved between Qur'an, Hadith, and Seerah resources, what kinds of trust or credibility judgments they made, and what features they would want in a more interconnected learning experience.

All interviews were conducted in English. All interviews were audio recorded and transcribed for analysis. Where needed, we retained familiar Islamic terms in transliterated or established English form rather than flattening them into generic paraphrase. Because our interest was in participants' accounts of practice, meaning, and design expectations, we treated the interview encounter as a site for eliciting situated reflection rather than as a standardised instrument \citep{kallio2016semistructured}.

\subsection{Data analysis and theoretical framing}

We analysed the interviews using reflexive thematic analysis \citep{braun2006thematic,braun2019reflecting}. We began with repeated reading of the transcripts to develop familiarity with participants' study routines, frustrations, trust practices, and design suggestions. We then conducted initial coding across the corpus, focusing on recurring concepts such as contextual understanding, authenticity, switching between apps, reflection, planning, study depth, and the tension between contextual expansion and reading flow. During this stage, we wrote analytic notes on emerging patterns, tensions, and cases that complicated initial interpretations. We compared and grouped these codes into candidate themes, reviewed them against the full dataset, and refined them into the final thematic structure reported in this paper. Theme names, boundaries, and supporting quotations were revisited iteratively across the research team during memoing and drafting. We retained five themes because they captured the recurrent patterns most relevant to our research questions while preserving meaningful differences between participants.

Our analysis was inductive in its initial coding, but three recurring dimensions consistently organised later memoing and theme development. First, participants described app use through situated routines and access conditions, which made gendered digital religion analytically useful. Second, participants judged apps and sources through authenticity, visible provenance, commentary, and the effort required to verify content, which made epistemic trust central. Third, they described learning as moving across phones, websites, laptops, and changing study intensities, which made seamless learning and cross-device continuity relevant. We therefore interpreted the themes through these three sensitising lenses rather than treating them as fixed a priori coding categories \citep{braun2019reflecting,lovheim2013media,nas2022women,fogg1999credibility,fogg2003prominence,corritore2005mechanics,kearney2020seamless,brudy2019crossdevice}. Consistent with reflexive thematic analysis, we did not treat coding as a reliability exercise or seek agreement between coders; instead, we treated theme development as an interpretive process grounded in memoing, rereading, and iterative comparison across the dataset \citep{braun2019reflecting,braun2021saturation}. This approach allowed us to remain close to participants' accounts while still engaging broader HCI concerns around credibility, situated practice, and interaction design across contexts \citep{crabtree2025human}.

\subsection{Author positionality and reflexivity}

This study examines religious learning practices in a culturally and spiritually sensitive domain. Following qualitative scholarship on reflexivity and positionality, we treat analysis as shaped by researchers' relation to the field rather than as independent from it \citep{england1994getting,finlay2002negotiating,berger2015now}. Our relationship to this domain is best described as partial insiderhood: we were familiar with Islamic learning vocabularies, commonly used digital Islamic resources, and the practical significance of questions around authenticity, commentary, and studying across sources, yet we still had to render those assumptions explicit for a broader HCI readership. This familiarity helped us recognise nuances that might otherwise have been missed, but it also risked making some patterns appear more obvious than they were. We therefore approached participants' accounts not as transparent confirmation of prior assumptions but as situated perspectives requiring careful interpretation.

Reflexivity was especially important because this was a study centred on Muslim women conducted through a community linked to the course and because the interviewer was female. In this context, using a female interviewer was part of how the study addressed access, trust, and disclosure in a participant group that is often underrepresented in HCI work on Muslim communities. These conditions likely shaped the kinds of examples participants chose to share, as well as how openly they discussed religious routines, app use, and design frustrations. Throughout analysis and writing, we therefore asked how our own familiarity with the field, and our decision to read the material through gendered digital religion, epistemic trust, and seamless learning, illuminated some aspects of the data while potentially muting others \citep{england1994getting,finlay2002negotiating,berger2015now}. We used these lenses as interpretive prompts rather than as exhaustive explanations.

\subsection{Ethics}

Ethical approval for this study was granted by Queen Mary University of London. All participants provided informed consent before taking part in the interviews. Because recruitment occurred within a learning network linked to the course, we treated confidentiality and voluntariness as especially important: participation was optional, we report participants pseudonymously, and we remove potentially identifying details from quotations and descriptions.

\section{Findings}

Across the interviews, we identified five themes that capture how participants understood and evaluated interconnected learning in Islamic apps.

\begin{table}[ht]
\centering
\caption{Summary of themes and central design tensions}
\label{tab:themes}
\small
\begin{tabularx}{\textwidth}{p{2.8cm}Y Y}
\toprule
Theme & Core analytic claim & Illustrative participant formulation \\
\midrule
Contextual and practical understanding & Participants wanted links between Qur'an, Hadith, and Seerah in order to build a coherent narrative, interpret difficult passages, and relate learning to daily life. & P2 described interconnected learning as giving a \enquote{holistic picture of Islam}; P4 said it offered \enquote{a more holistic context to the verse}. \\
Trust and authenticity & Integration mattered when it reduced verification work and presented authentic sources with explanatory commentary. & P1 valued a source she could \enquote{trust with our eyes closed}; P5 said, \enquote{I don't trust Google}. \\
Seamless integration without clutter & Users wanted one place with optional depth, summaries, and expansion within the same view rather than being forced to move between apps or navigate overloaded interfaces. & P4 wanted \enquote{everything in one place... pop ups rather than... taking me to a different place}; P3 warned against features that \enquote{destroy the flow}. \\
Different modes of use & Mobile apps were useful for devotional continuity and routines, but deeper study often shifted to laptops, websites, search, and note taking workflows. & P4 said, \enquote{Since I live in the UK, I don't have regular adhan. So I use the Pillars app}; P5 stated that \enquote{for study, I do tend to prefer using my laptop}. \\
Guidance and habit support & Participants wanted apps to scaffold practice through planners, quizzes, reflection prompts, and progress tracking. & P2 proposed a \enquote{personal goal board}; P5 wanted reflective questions because \enquote{when you write something, you just give it more thought}. \\
\bottomrule
\end{tabularx}
\end{table}

\subsection{Theme 1: Interconnected learning is about contextual and practical understanding}

We found that participants consistently described the value of connected Qur'an, Hadith, and Seerah content in terms of context rather than convenience alone. They wanted to know where a verse came from, what was happening in the Prophet's life, and how a teaching should be understood in practice. P2 described this as a \enquote{holistic picture of Islam rather than just, like, reading a verse from the Qur'an}. P4 likewise said that seeing Qur'an, Hadith, and Seerah \enquote{side by side... gives a more holistic context to the verse}.

For P3, the relationship between these sources was constitutive rather than optional: \enquote{you can't really understand the Qur'an if you do not understand our prophet's life, so the Seerah. And then you cannot necessarily follow the Hadith if you don't know the Qur'an}. Participants thus framed interconnection as interpretive scaffolding. It helped them avoid isolated reading, connect teachings to prophetic biography, and see how scripture applies in everyday life.

This was especially clear when participants discussed sensitive or contested topics. P3 explained that certain issues become easier to understand once the app situates them historically and interpretively: \enquote{Just one sentence doesn't necessarily mean a lot unless you can put it in a particular situation}. P4 made a similar point more directly: \enquote{Sometimes we misinterpret certain verses, we just look at the literal wording, and that's where our knowledge is very limited}. In this sense, interconnected learning was not imagined as extra information, but as a way of preventing partial or distorted understanding.

\subsection{Theme 2: Trust and authenticity are central to why integration matters}

We found that participants did not want integrated interfaces simply because they are efficient. They wanted them because trusted integration reduces the need to verify, compare, and check across multiple sources. This was most explicit in P1's account. P1 said, \enquote{We can trust with our eyes closed that the tafsir, duas, and hadith here are authentic}. The same participant connected authenticity directly to reduced labour: \enquote{This is a major support for us because we do not have to keep searching around in different places}.

The interviews reinforced this concern from another angle. P5 described using search engines to find hadith quickly, but immediately qualified that practice: \enquote{I don't trust Google... and then I'll go... to check it... just to make sure}. In other words, convenience without trust is insufficient. Several participants also stressed that mere linking is not enough; the surrounding commentary is what makes the connection useful. P5 put this plainly: \enquote{if that page when you go to the Hadith has a commentary, then, obviously, that's super helpful}.

This theme suggests that trust in Islamic learning apps has at least two components. The first is source trust: confidence that content is authentic, well curated, and responsibly presented. The second is interpretive trust: confidence that the app helps users understand why a source matters, rather than merely pointing to it. Integrated design, then, is partly an epistemic design problem.

\subsection{Theme 3: Users want seamless integration without clutter or interruption}

We found that participants wanted interconnected learning to feel seamless, but not intrusive. The dominant preference was for one integrated environment rather than multiple apps or disruptive jumps across interfaces. P2 wanted \enquote{integrating everything in one app... making it look seamless rather than, like, some links}. P4 similarly said, \enquote{I would like everything in one place... pop ups rather than... taking me to, like, a different place where I might just get lost}. P1 made a similar point, arguing that people do not want to go somewhere else to see or do something; if everything connected to a verse is available in one place, they will read it there.

At the same time, participants did not want all contextual material forced into the main flow. P3 warned that if Seerah reading were interrupted by frequent hadith inserts, it would \enquote{destroy the flow of the Seerah}. P5 preferred \enquote{a summary of the story with the reference in bracket and linked}, arguing that the full text of a long hadith does not always read well on screen. These comments point toward layered disclosure: keep the primary reading experience intact, but make deeper context immediately available when desired.

The strongest cautionary note came from P4, who said that when \enquote{one app tries to include too many features, it becomes a bit overwhelming. So it has to be simple to use}. The same participant also complained that \enquote{ads get really annoying} and should not \enquote{disrupt the learning experience}. Taken together, these comments show that successful interconnected design must increase coherence without increasing friction.

\subsection{Theme 4: Interconnected learning must support different modes of use}

The interviews revealed two distinct, though overlapping, usage modes. The first was lightweight devotional or routine focused mobile use: reading while travelling, listening during periods when recitation practices change, checking prayer times, reading duas, or making use of bookmarks and \enquote{last read} states. P2 described app use as helpful precisely because \enquote{we are always busy} and carrying a phone is easier than carrying hard copies. P4 likewise relied on a prayer app for accurate timings and routine planning, explaining, \enquote{Since I live in the UK, I don't have regular adhan. So I use the Pillars app.} She added that the app helped with time management and regularity in prayer. P5 found the Qur'an app particularly useful when listening to Arabic recitation alongside an English translation on her phone.

The second mode was deeper study or research work. Here, mobile apps were often seen as secondary. P5, who described herself as doing serious Islamic study, said plainly: \enquote{for study, I do tend to prefer using my laptop}. She explained that search, Arabic commentaries, copying and pasting, and document writing are central to her workflow, and therefore she needs web or desktop access: \enquote{being able to copy paste it is, like, really, really important}. She also noted that if interconnected content existed only inside a mobile app, it would not fit how she actually studies.

These accounts complicate any assumption that a better mobile app alone will solve the design problem. The data suggest a broader learning ecology in which mobile apps support devotional continuity and lightweight access, while websites or desktop tools support research, writing, and deeper comparison. Designing interconnected learning may therefore require thinking beyond a single device class.

\subsection{Theme 5: Users want the app to guide learning, not merely store content}

We found that participants repeatedly proposed features that move beyond content access towards learning support and habit formation. Planning tools were especially salient. P2 wanted a \enquote{personal goal board} showing how much had been read. P3 suggested an Umrah itinerary and described planners as useful because they provide direction and reduce distraction. P1 asked for a memorisation planner that could break goals into manageable parts and help users track progress over time.

Quizzes and reflective prompts formed a second cluster of requests. P4 said quizzes would make the experience more interactive and \enquote{keep me motivated}. P5 wanted reflective questions attached to readings, and emphasised that writing down a response changes the quality of engagement: \enquote{When you write something, you just give it more thought}. She also proposed a \enquote{pause and reflect} pattern that could nudge users from recitation towards contemplation and application.

These ideas show that users do not see an interconnected app as a static repository. They imagine it as a companion for practice, memorisation, reflection, accountability, and study continuity. The interface is therefore expected not only to connect sources, but also to guide learning behaviour over time.

\section{Discussion}

\begin{figure}[ht]
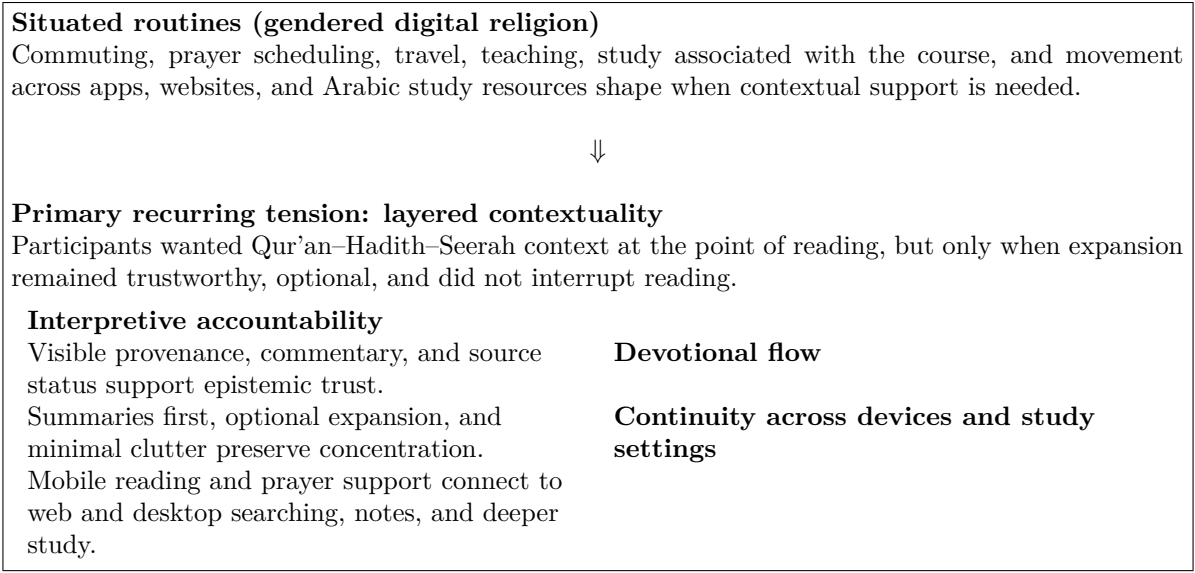

\centering
\fbox{\parbox{0.94\textwidth}{
\small
\textbf{Situated routines (gendered digital religion)}\\
Commuting, prayer scheduling, travel, teaching, study associated with the course, and movement across apps, websites, and Arabic study resources shape when contextual support is needed.

\begin{center}
$\Downarrow$
\end{center}

\textbf{Primary recurring tension: layered contextuality}\\
Participants wanted Qur'an--Hadith--Seerah context at the point of reading, but only when expansion remained trustworthy, optional, and did not interrupt reading.

\vspace{0.5em}
\begin{tabularx}{\linewidth}{>{\raggedright\arraybackslash}X >{\raggedright\arraybackslash}X >{\raggedright\arraybackslash}X}
\textbf{Interpretive accountability}\\
Visible provenance, commentary, and source status support epistemic trust.
&
\textbf{Devotional flow}\\
Summaries first, optional expansion, and minimal clutter preserve concentration.
&
\textbf{Continuity across devices and study settings}\\
Mobile reading and prayer support connect to web and desktop searching, notes, and deeper study.
\end{tabularx}
}}
\caption{Layered contextuality in interconnected Islamic learning. The model summarises the recurring tension across the interviews and its main design dimensions.}
\label{fig:layered-contextuality}
\end{figure}

The findings identify layered contextuality as the central design issue in interconnected Islamic learning. Across the interviews, participants consistently sought Qur'an--Hadith--Seerah context at the point of reading, but only in forms that remained inspectable, optional, and compatible with sustained reading. Interpreted through the three sensitising lenses, this pattern locates the design problem in situated study practices rather than in an abstract preference for feature integration. Figure~\ref{fig:layered-contextuality} summarises this account.

\subsection{Layered contextuality in interconnected Islamic learning}

Existing work shows that Muslim users value trust, flexibility, and spiritual relevance in Islamic apps, while reviews of Qur'an learning tools note the prevalence of discrete functions over integrated interpretive support \citep{mubin2020reviewing,kabir2025islamiclifestyle}. The present findings extend this literature by showing that participants repeatedly sought context anchored to the relation among Qur'an, hadith, and Seerah at moments of interpretive difficulty, and that they wanted such context to preserve reading flow rather than replace it with extended scholarly exposition.

The analytical value of a sample centred on Muslim women becomes clearer in this context. As the findings indicate, participants linked app use to specific routines: relying on prayer apps in the UK where regular adhan was not available, travelling, teaching, study tied to the course, and movement between mobile apps, websites, and Arabic study resources. Read through the lens of gendered digital religion, these routines are not incidental background details but the conditions through which the design problem becomes visible \citep{lovheim2013media,nas2022women,nisa2022muslims,bardzell2010feminist,bardzell2011feminist}. Although the cohort is not treated as representative of Muslim women as a whole, it demonstrates how an underrepresented user group can recast an HCI problem from content aggregation to the more specific question of how contextual links, source explanation, and study continuity should be staged in practice.

Islamic scholarship further situates this pattern without overstating the scope of the findings. Contextualist work in Qur'anic studies and hadith scholarship on \enquote{asb\={a}b al-wur\={u}d} both treat text, circumstance, and explanatory tradition as interdependent \citep{saeed2021contextualist,calis2022contextual,ramle2022between}. Participants were not engaged in formal scholarly method, but their interface requests align with a pedagogical logic already recognised within Islamic scholarship: meaning often becomes clearer when verse, report, narrative setting, and explanation can be read together. From an HCI perspective, layered contextuality therefore refers to more than linking across sources; it concerns the staging of interpretive support so that context becomes available when needed rather than being imposed in advance.

\subsection{Interpretive accountability}

Participants did not assess trust solely through the general credibility of an app. They were equally concerned with whether connections among sources were scholastically accountable. They wanted to know what collection a hadith came from, whether commentary was present, which interpretive voice was being surfaced, and why a linked source was relevant. The term interpretive accountability captures this dimension of trust: confidence that linked material is sourced, explained, and situated clearly enough for users to judge its religious authority.

Research on credibility in HCI shows that users assess trust through information quality, presentation, and interface cues \citep{fogg1999credibility,fogg2003prominence}. Riegelsberger et al. further argue that trust in human--computer interaction is relational as well as perceptual \citep{corritore2005mechanics}. The present findings indicate that, in Islamic learning systems, credibility is also shaped by whether interfaces make provenance, commentary, and interpretive status inspectable. This interpretation resonates with digital religion scholarship showing that online Islamic authority is negotiated rather than automatically granted \citep{campbell2013digital,bunt2009imuslims,solahudin2019internet}, and with participants' repeated resistance to decontextualised search results or anonymous explanatory text. In this study, trust was inseparable from the ways interfaces exposed the lineage and status of religious knowledge.

\subsection{Continuity across devices and study settings}

The seamless learning lens foregrounds another recurring pattern: participants wanted contextual support within an ecology of devices and study intensities rather than within a single integrated app. Mobile apps were valued for portability, recitation, prayer support, and quick reference, whereas deeper study often shifted to websites, laptops, workflows built around copying and pasting, and research centred on note taking. This pattern aligns with cross-device HCI and seamless learning research, while also indicating that, in religious learning, continuity is tied not only to productivity but also to devotional rhythm, spiritual focus, and movement between quick use and deeper interpretive work \citep{kearney2020seamless,brudy2019crossdevice}. The findings therefore refine earlier Islamic app research by showing that flexibility includes movement between devotional use, contextual lookup, commentary checking, and sustained study rather than merely personalised settings or reminders \citep{kabir2025islamiclifestyle}.

Participants' proposed guidance features can be interpreted through the same lens. Planners, itineraries, memorisation aids, quizzes, and reflective prompts position the app less as a repository of content and more as an optional companion for reflective practice, provided such supports remain proportionate to the task \citep{sengers2006reflective,wong2019selfregulated,kori2014reflection,markum2023religious}. The design implications that follow translate these interview patterns into a set of HCI considerations centred on layered contextuality, interpretive accountability, and continuity across devices and study settings.

\section{Design Implications}

Table~\ref{tab:implications} translates recurring tensions in the interviews into design implications situated in adjacent HCI and Islamic scholarship \citep{mustafa2020islamichci,hakkila2020cultural,alshehri2022culturally}. As the study examined articulated needs rather than implemented systems, these implications should be understood as empirically grounded directions for design rather than as evaluations of specific solutions. The first three implications address the core problem of layered contextuality by specifying how context should be introduced, explained, and staged. The final two address continuity across study settings and the role of optional scaffolds in longer term engagement.

\begin{table}[!htbp]
\centering
\caption{Design implications grounded in the study and related literature}
\label{tab:implications}
\small
\begin{tabularx}{\textwidth}{p{2.7cm}Y Y}
\toprule
Design implication & Grounding in the interviews & HCI/system response \\
\midrule
Anchor context to the reading unit & Participants wanted help at the moment a verse, passage, or story became difficult to interpret; they preferred concise, relevant contextualisation over undifferentiated repositories. & Link hadith, Seerah episodes, or commentary to the current reading unit and briefly state why the link is relevant. This responds to fragmentation in Islamic app ecologies and aligns with contextualist scholarship \citep{mubin2020reviewing,saeed2021contextualist,calis2022contextual,ramle2022between}. \\
Make provenance inspectable & Trust depended on authenticity, commentary, and the effort required to verify a source. Participants wanted to see why a connection could be trusted, not just that it existed. & Expose collection or source, grading or status where appropriate, translator or commentator, and whether explanatory text is editorial, scholarly, or generated by users. This extends credibility research toward interpretive accountability \citep{fogg1999credibility,fogg2003prominence,corritore2005mechanics,solahudin2019internet}. \\
Use optional layered disclosure & Participants wanted more context without clutter, redirection, or a reading experience that felt like reading a thesis. Several explicitly preferred summaries plus expandable references. & Favor bracketed references, expandable cards, bottom sheets, or short inline summaries over forced external links. The design goal is to preserve devotional flow while making depth available on demand \citep{okuboyejo2021concerns,sengers2006reflective}. \\
Support continuity across devices and study settings & Mobile apps supported recitation, prayer routines, and quick lookup, while deeper study often moved to laptops, web search, taking notes, and workflows built around copying and pasting. & Synchronise reading state, bookmarks, notes, and saved references across mobile and web, and support exporting material into workflows focused on study. This follows seamless learning and cross-device HCI \citep{kearney2020seamless,brudy2019crossdevice,kabir2025islamiclifestyle}. \\
Offer optional scaffolds for specific scenarios & Desired support features were tied to concrete practices such as Umrah preparation, memorisation, recitation goals, and reflective writing rather than generic productivity. & Make planners, prompts, quizzes, and memorisation aids tied to specific scenarios and keep them optional so that they guide without coercing. This aligns with self-regulated and reflective learning research while staying close to participants' stated routines \citep{wong2019selfregulated,kori2014reflection,markum2023religious}. \\
\bottomrule
\end{tabularx}
\end{table}

Taken together, these implications point to a design orientation that is more specific than a general call for better integration. In this study, an effective interconnected system would not simply aggregate more Islamic content within a single interface. It would introduce context at the point of need, make the sourcing of that context visible, preserve reading flow, and support movement between quick devotional use and deeper study.

\FloatBarrier

\section{Limitations and Future Work}

This study draws on five Muslim women from a single online Islamic learning community. Its analytical value lies in the specificity of that cohort: it provides a focused account of how an underexamined participant group articulated requirements for interconnected Islamic learning, and that specificity is consistent with information power as a rationale for focused qualitative inquiry \citep{malterud2016information,crabtree2025human}. At the same time, the findings are not intended as statistically generalisable claims, and some patterns may reflect the pedagogical norms, shared vocabulary, and study commitments of this particular course-linked community. The strength of the study therefore lies in the depth and relevance of a bounded cohort rather than in broad transferability across Muslim populations.

The sample was also socially, educationally, and infrastructurally specific. All participants were women in their 30s and 40s, highly educated, digitally engaged, connected to an Islamic studies course environment, and Sunni Muslims. Employment status was similarly narrow in aggregate, with one housewife, one full-time student, and three participants in full-time employment. Recruitment through a Muslim women's WhatsApp group associated with a structured course likely privileged women who were already comfortable with organised religious learning, smartphone use, and participation in networked learning communities. As a result, the study may underrepresent Muslim women whose access to digital technologies, religious learning spaces, or supportive social infrastructures is more constrained, an issue that prior HCI and Islamic technology research has already shown to be unevenly distributed \citep{mustafa2020islamichci,ibtasam2019family,hardaker2017perceptions}. The study therefore does not claim to represent Muslim women app users more broadly, or Muslim users in general, and it did not examine denominational or occupational variation as an analytic dimension.

The study was also shaped by language and modality choices. All interviews were conducted online and in English, which likely favoured participants who were comfortable with remote interviewing, extended verbal reflection, and discussing Islamic learning practices in English. In practice, this aligns with the profile of the sample itself, which was predominantly UK-based, postgraduate educated, and already accustomed to digital Islamic resources. These choices were appropriate for the present study, but they also mean that the accounts captured here are most reflective of women who could participate under those linguistic and technical conditions rather than of Muslim women whose access, confidence, or preferred language might differ \citep{mustafa2020islamichci,ibtasam2019family,hardaker2017perceptions}.

Methodologically, the study is based on semi-structured interviews rather than observed use, diary studies, or prototype evaluation. It therefore captures articulated needs, retrospective judgments, and desired features, but not observed behaviour, interactional breakdowns, or in-the-moment interpretive decisions during actual app use \citep{kallio2016semistructured,crabtree2025human}. The themes should also be understood as interpretive outcomes shaped by reflexive thematic analysis, by the research team's partial insider position, and by the decision to read the material through gendered digital religion, epistemic trust, and seamless learning \citep{braun2019reflecting,england1994getting,finlay2002negotiating,berger2015now}. Alternative analytical framings might reasonably have foregrounded other dimensions, including denominational practice, platform politics, or pedagogical hierarchy. Future work should therefore extend this study through more varied Muslim samples across learning backgrounds, access conditions, and interpretive communities, including participants less closely tied to formal course structures, and through methods that combine interviews with diary, observational, and prototype-based inquiry. Such work would be well placed to test contextual links, provenance cues, reflective prompts, planners, and cross-device continuity in situated use.

\section{Conclusion}

This paper examined how Muslim women in a learning community linked to a course understood interconnected learning in Islamic apps. The findings show that layered contextuality is central to this design problem: participants wanted richer Qur'an--Hadith--Seerah context, but only in forms that preserved interpretive accountability, devotional flow, and continuity across different study intensities.

For HCI, the study clarifies how contextual expansion, provenance, and continuity across devices and study settings converge in a domain where trust depends on religious interpretation and source accountability, not only on presentational credibility cues. Viewed through the lenses of gendered digital religion, epistemic trust, and seamless learning \citep{lovheim2013media,nas2022women,fogg1999credibility,corritore2005mechanics,kearney2020seamless,brudy2019crossdevice}, these findings establish a concrete agenda for future work: participatory studies based on prototypes that test how contextual Islamic scholarship can be represented in ways that remain trustworthy, readable, and spiritually appropriate in practice.

\section*{Disclosure Statement}
The authors report there are no competing interests to declare.

\section*{Funding}
No funding was obtained for the reported work.

\section*{Data Availability Statement}
The interview recordings and transcripts are not publicly available because they contain identifiable religious learning experiences collected under confidentiality and anonymisation procedures.

\section*{Declaration of Generative AI Use}
The authors used generative AI for manuscript drafting and language editing assistance. Generative AI was not used for data collection, transcription, coding, analysis, interpretation, or the generation of findings.

\bibliographystyle{apalike}
\bibliography{ijhci_interconnected_learning}

\end{document}